\documentclass{elsart}

\usepackage{natbib}

\usepackage{amssymb}

\usepackage[pdftex]{graphicx}
\usepackage{url}
\usepackage{color}

\definecolor{hcyan}{rgb}{0,0.66,0.92}
\definecolor{hblue}{rgb}{0.16,0.15,0.52}
\definecolor{hpink}{rgb}{0.87,0.06,0.5}
\definecolor{hred}{rgb}{0.87,0.06,0.15}
\definecolor{hgreen}{rgb}{0.,0.62,0.25}
\definecolor{hyellow}{rgb}{0.97,0.92,0.18}

\begin{document}

\begin{frontmatter}



\title{Incorporating interactive 3-dimensional graphics in astronomy research papers}
%\title{Interactive visualisation in the digital publishing era}


\author{David G.\ Barnes\corauthref{cor1}} and
\corauth[cor1]{Corresponding author.}
\ead{dbarnes@swin.edu.au}
\author{Christopher J.\ Fluke}
\ead{cfluke@swin.edu.au}

\address{Centre for Astrophysics and Supercomputing, 
Swinburne University of Technology, 
PO Box 218, Hawthorn, Victoria, Australia, 3122}

\begin{abstract}
Most research data collections created or used by astronomers are
intrinsically multi-dimensional.  In contrast, all visual
representations of data presented within research papers 
are exclusively 2-dimensional.  We
present a resolution of this dichotomy that uses a novel technique for
embedding 3-dimensional (3-d) visualisations of astronomy data sets in
electronic-format research papers.  Our technique uses the latest
Adobe Portable Document Format extensions together with a new version
of the S2PLOT programming library.  The 3-d models can be easily
rotated and explored by the reader and, in some cases, modified.  We
demonstrate example applications of this technique including: 3-d
figures exhibiting subtle structure in redshift catalogues,
colour-magnitude diagrams and halo merger trees; 3-d isosurface and
volume renderings of cosmological simulations; and 3-d models of
instructional diagrams and instrument designs.
\end{abstract}


\end{frontmatter}

% main text
\section{Introduction}
\label{}

%\begin{quote}
%{\em Editor's note: the electronic version of this paper exhibits the new publishing
%features discussed herein.  We therefore encourage readers to examine
%this paper in its electronic form, rather than in print.}
%\end{quote}

\subsection{Visualisation in astronomy}

% astro datasets are big and need visualisation
Astrophysical data collections (datasets) are predominantly large and
multi-dimensional.  Accordingly, the principle approach to studying
and comparing datasets is to calculate statistical measures such as
the moments (mean, skewness, etc.) and correlation functions.  Even
substantial techniques such as Fourier analysis and principal
components analysis rely heavily on basic statistical measures.  These
methods provide valuable information on the ensemble properties of the
datasets.  What they are less well-suited to is identifying
anomalies or special cases within the data: are there features that we
cannot understand or explain based on purely statistical approaches?
Usually, there is a great deal of information that can be obtained
simply by {\em looking} at the data.  This is called scientific
visualisation, hereafter simply ``visualisation''.

% categories of visualisation
Broadly speaking, visualisation methods can be classified as being
qualitative (direct visual inspection), quantitative (including
selection of regions of interest and computation of statistical
properties), or comparative (including comparisons between objects
with secondary observational catalogues, side-by-side comparisons
between datasets, or overlays such as isodensity surfaces).

% early history of visualisation
Visualisation in various forms has played a role in astronomy since
early times.  The Ancient Egyptians mapped the locations of prominent
stars in their tomb decorations, Greek constellations are seen
inscribed on the Farnese globe (c. 2nd century CE) and Chinese star
maps have been found dating back to 940 CE.  The 14th and 15th
centuries saw the development of star atlases of increasing accuracy
and beauty - truly works of ``cognitive art'' \citep{Tufte90} in their
own right. \citet{Beniger78} and the extremely comprehensive web-site
of Friendly \& Denis\footnote{Milestones in the History of Thematic
Cartography, Statistical Graphics, and Data Visualization:
http://www.math.yorku.ca/SCS/Gallery/milestone} cover the historical
development of graphic representations (with passing references to
their use in astronomy), including tables, coordinate systems and
maps, and the derived forms: line graphs, histograms, scatter plots and
so on.

% computer graphics revolution
The advent of computer graphics revolutionised visualisation in
astronomy.  It enabled dynamic presentation, and more importantly,
real-time exploration of astrophysical datasets.  In general, an
interactive digital representation allows the user to actively rotate,
zoom, and pan multi-dimensional datasets in order to find a viewpoint
that provides more information and understanding.  This is {\em
visualisation-led knowledge discovery} and is now a standard and
necessary tool in all of the physical sciences.

The first computerised astrophysical visualisation systems were
expensive to operate and difficult to use.  For example, the
3-dimensional representations of the Centre for Astrophysics (CfA)
galaxy surveys, which helped reveal the ``Great Wall'' and other major
filamentary features \citep[e.g.][]{Geller89}, required custom
processors.  Image sequences were generated by outputting individual
frames that were then printed to 16 mm film for viewing!  Real-time
interactive motion was merely ``a fond hope''
\citep{Geller92}.

%{\color{red} Chris to add further examples here if necessary.}

Today, an entry-level laptop or workstation provides an extraordinary
level of graphics performance.  In fact, in many modern desktop
computers, the graphics processing unit is actually more powerful than
the central processing unit.  Consequently, there is now a wide
selection of commercial and free software applications, programming
libraries and data processing environments that provide varying levels
of 3-dimensional graphics capabilities to scientists.  It is fair to
say that astronomers now can, and increasingly {\em do}, make use of
advanced 3-dimensional visualisation in their quest for knowledge
discovery.

%, yet this
%processing power is currently under-utilised for scientific computing.
%Recent advances in computer graphics hardware and display technologies
%(e.g. large format LCD panels, plasma screens and large-format digital
%projectors) have been driven by the ever-growing computer games and
%home entertainment markets.  This has enabled the development of
%low-cost advanced image displays, such as stereoscopic environments
%(using passive polarizing, electronic shutter glasses, or glasses-free
%autostereoscopic panels) and digital domes, however, the use of such
%displays is not widespread amongst astronomers \citep{Fluke06}.

\subsection{Publishing in astronomy}

% www has changed publishing
The growth of the World Wide Web \citep{BernersLee94} as a global
information repository and communication tool has profoundly affected
the way science is performed and reported.  In particular, there has
been a dramatic change in how research articles are published, with a
steady trend away from physical, paper-based journals to fully online
digital publications.  The way was paved in the early 1990s, as the
Astrophysical Journal (ApJ) introduced video tapes for ``illustrations
that are not well suited to single frames or figures'' \citep{Abt92}
and CD-ROMS of datasets, both of which were distributed with the paper
editions.  The first online edition of ApJ appeared in September 1995
\citep{Abt02}.  During the same period, the ADS Abstract Service began its 
service, providing astronomers with new capabilities to search for and
obtain abstracts \citep{Murray92,Kurtz00}, followed soon
after with the provision of scanned articles \citep{Eichhorn94}.
%  The
%astronomical community enthusiastically joined the digital publishing
%revolution, and the number of visits to campus or observatory
%libraries to access paper-based journals rapidly decreased.

% yet figures remain restricted to 2-dimensional representation
Today, the standard graphics cards that ship with new computers are
immensely powerful.  They are perfectly suited to running the
advanced, 3-d visualisation programs that are becoming necessary for
the rapid comprehension of large and complex data collections.  Yet
despite the nearly complete migration to an electronic workflow in
research publishing, published visualisations remain almost
exclusively {\em two-dimensional}\footnote{Stereoscopic images can be
printed, usually to be viewed with coloured anaglyph or
chromastereoscopic glasses, but their use in research publications has
been limited.} and {\em static}.  The former embodies a graphical
communication challenge, as the majority of research data collections
in astronomy are multi-dimensional and can in many cases be
communicated better with 3-d presentation.  The latter limits the
presentation of time-evolving data to ``cartoon strips'' of individual
frames.  While the major journal publishers have allowed {\em ex
situ}\/ supplementary material to be linked from published papers, the
{\em in situ} publication of 3-dimensional, interactive figures
remains largely unexplored.

%\footnote{The principle exception to this are
%the elaborate volvelles or rotula that were introduced in the 13th
%century by English historian Matthew Paris.  A volvelle uses a
%combination of rotatable paper disks and string markers, which were
%bound into astronomy texts as early interactive displays.  Now mainly
%a historical curiousity, volvelles have evolved into the modern
%cardboard or plastic planispheres.  For discussions of the use of
%volvelles in astronomy, ranging from textbooks to medical diagnosis
%via astrology, see \citet{Bruck98}, \citet{Gingerich90},
%\citet{Robbins70} and \citet{Stebbins59}.}.  

\subsection{This paper}

The latest extensions to the Adobe Portable Document Format (PDF)
create an exciting opportunity to bring together the interactivity of
a real-time, 3-dimensional visualisation system with the requirements
for standardised academic publication (within an electronic document
context).  In this paper, we describe a relatively straightforward
technique for developing and embedding interactive, 3-dimensional
science visualisations as figures in PDF files that can be viewed 
with the freely downloadable Adobe Reader application.  We use the S2PLOT library
\citep{Barnes06} and Adobe Acrobat 3D Version 8.1 software, although
other approaches are possible (Goodman et al., {\em priv. comm.}).

In Section~\ref{sec:3dpdf} we introduce Adobe's 3-d extensions to PDF
that our technique exploits and motivate the use of 3-dimensional PDF
for science reporting.  In Section~\ref{sec:s2plot} we review S2PLOT
as a tool for generating 3-dimensional, scientific models, and
describe how these models can be saved and imported for embedding in
existing or new PDF files.  Advanced, interactive features possible
with JavaScript are discussed in Section~\ref{sec:js}, and alternative
uses of 3-d PDF are described in Section~\ref{sec:other}.  Examples
are given throughout the paper.  We conclude the paper in
Section~\ref{sec:conc} with some brief commentary on the technique we
present, and a few remarks on future applications of 3-d figures in
astronomy papers.

\section{3-dimensional PDF}
\label{sec:3dpdf}

\subsection{Portable Document Format}

Portable Document Format (PDF) is an open document standard developed
by Adobe Systems Incorporated.\footnote{Adobe Systems Inc.:
\url{http://www.adobe.com}}  First released in the early 1990s, PDF
has  become the preferred standard for printable and 
and electronically-distributed documents,  supplanting its predecessor 
PostScript -- Adobe's first document standard.  PDF files may
contain text, vector graphics and bitmap graphics, and can be produced
by most commercial desktop publishing software systems, and by many
free applications such as the \LaTeX\ typesetting system.

Uptake of PDF by journal publishers is virtually complete in the sense
that nearly all publishers have adopted PDF as a standard target file
type, for both pre-press operations and electronic distribution.
Astronomy has been no exception, with the major journals using PDF for
most parts of the paper publication and distribution workflow since
the late 1990s.  The ability of \LaTeX\ to produce PDF files,
initially via conversion from PostScript and then directly, was
pivotal in the embracing of PDF by the physical sciences research
community.

\subsection{Acrobat 3D and Adobe Reader}

In May 2007, Adobe announced the availability of ``Adobe Acrobat 3D
Version 8'' software.\footnote{Acrobat 3D Version 8 announcement:
\url{http://www.adobe.com/aboutadobe/pressroom/pressreleases/pdfs/200705/053007Acrobat3DShips.pdf}}
 This product builds on extensions to PDF which allow the inclusion of
 3-dimensional objects described in the Universal 3D (U3D) format.
 Acrobat Reader has been able to interactively display U3D objects in
 PDF files since version 7.  However, the release of Acrobat 3D
 Version 8 and the free Acrobat Reader Version 8 have standardised and
 dramatically simplified the creation and viewing of 3-d PDF files.

Acrobat 3D Version 8 (hereafter ``Acrobat 3D'') is directed at
Computer Aided Design (CAD) and Computer Aided Modelling (CAM)
professionals.  It can import 3-d models from more than 50 third-party
file formats.  Roughly two-thirds of the supported formats are
polygonal meshes from CAD/CAM applications, while the remaining formats cover
animation packages, game models, and more generic 3-d formats.  In
some circumstances, Acrobat 3D can also capture 3-d geometry from the
hardware graphics pipeline as it is displayed.

Once a 3-d model has been imported to Acrobat 3D, the publisher can
set the preferred lighting, shading and viewing for the model.  A
number of different configurations can be saved as views.  The model
tree can be exposed, and its various branches (corresponding to
particular geometric elements of the scene) can be switched in or out
of the visualisation.  Options on import allow colouring of different
parts of the model, and optimisation of properties such as surface
smoothness and detail.  It is evident from the example PDF files
supplied with Acrobat 3D that Adobe's focus is firmly on the ability
to embed and share 3-d engineering models in PDF files.

Acrobat 3D also supports the addition of JavaScript to a PDF file, which
can be used to programmatically control almost every aspect of a 3-d
model visualisation.  This control can be intrinsic (ie.\ JavaScript
event handlers that respond implicitly to user interaction with the
3-d model), or extrinsic.  The latter entails embedding traditional
graphical user interface (GUI) elements into the PDF document and
modifying the 3-d visualisation in response to explicit user
input via those elements.  

With Adobe Reader Version 8 (hereafter ``Reader'', freely available
from Adobe), 3-d models embedded in PDF files are interactively
viewable.  All properties of the model set in Acrobat 3D are expressed
in the Reader, including saved views.  Unless explicitly disabled in
Acrobat 3D, Reader allows the user to interactively rotate the model,
select their own shading and lighting options, and explore the model
tree.  And importantly, JavaScript control of the embedded 3-d models 
extends to the Reader.

%The new publishing recipe? some heading?
%
%Consider a traditional PDF file containing narrative text and vector
%or bitmap graphics.  Add one or more embedded 3-d models, a few GUI
%elements (buttons, checkboxes, etc.), and some relatively simple
%JavaScript code.  The result is a self-contained document which, when
%viewed in Reader, constitutes an application of sorts.  The user has
%access to all the customary content of a text article, {\em plus}\/
%interactive, visualisations of 3-d objects.

\subsection{3-dimensional PDF for astronomy}

3-dimensional PDF holds great promise for improving science reporting,
We propose the judicious use of 3-d PDF to add 3-dimensional,
interactive figures to astronomy journal papers.  We contend that
there are many circumstances where the use of 3-d figures can be
substantially more illustrative of concepts, relationships and
properties, than can their 2-d counterparts.  While 3-d figures (and
more commonly, dynamic content such as movies) {\em have}\/ been
attached to astronomy papers \citep[for some recent examples
see][]{diemand07,okamoto07,price07} this has until now been achieved
using ephemeral web addresses to direct readers to supplementary
material.  Directly embedding 3-d figures in a standard document
format --- for which viewers exist on all major desktop platforms ---
affords a major improvement for both present useability and future
compatibility.

Despite Acrobat 3D being openly targeted at the CAD/CAM user
community, we show in this paper that 3-d PDF is also suitable for
scientific data.  Together with software that produces 3-d
models, Acrobat 3D can be used to produce publication-quality,
scientifically instructive 3-dimensional figures.  The basic geometric
elements required to produce scientific plots --- lines and points ---
are readily available, together with an extensive set of higher-level
objects such as surfaces and textures.

%When a PDF file with embedded
%3-d figures is printed, an author-selected static view of each 3-d
%figure is used.

\section{S2PLOT and 3-d PDF}
\label{sec:s2plot}

We now describe one approach to producing 3-d figures in PDF files,
using the S2PLOT programming library and Adobe Acrobat 3D Version 8.1.
S2PLOT \citep{Barnes06} is an advanced graphics library with a
programming interface familiar to users of PGPLOT.\footnote{PGPLOT is
written by T.J.\ Pearson:
\url{http://www.astro.caltech.edu/~tjp/pgplot/}.}  S2PLOT enables the
programmatic construction, display and interactive exploration of
3-dimensional scientific plots and diagrams.  S2PLOT can be called
from C, C++ and FORTRAN code, and is freely available for GNU Linux,
Apple OS X and Windows XP systems.\footnote{Windows XP support for
S2PLOT is provided via the Cygwin system:
\url{http://www.cygwin.com}.}  S2PLOT uses the OpenGL graphics
standard to exploit hardware-accelerated graphics performance, and
supports standard display devices such as desktop monitors and data
projectors, and advanced devices such as passive and active
stereoscopic systems and digital dome projectors \citep{Fluke06}.

S2PLOT is ideally-structured to produce 3-d model
output that can be imported into Acrobat 3D.  Internally, the S2PLOT
library maintains a list of all geometry that is used to make up a
scene. On each screen redraw, the active geometry is sent to the
graphics pipeline with standard OpenGL calls.  In S2PLOT version
2.0 the geometry list can be exported to a file in Virtual Reality Modeling
Language (VRML) format.  As well as being a supported input format for Acrobat
3D, this language has the distinct advantages that it is (i)
text-based, enabling easy editing prior to Acrobat 3D import; and (ii) web-based,
yielding further possibilities for model and figure sharing and
publication.  VRML has received occasional attention from
the astronomy community \citep[see e.g.][]{Plante99,Crutcher98},
and one of us (Barnes) has previously developed a VRML viewer for
Virtual Observatory data \citep{Beeson04}.

For basic 3-dimensional figures and diagrams, embedding an S2PLOT-based 3-d 
figure in a PDF paper is accomplished via a standard key press to export 
S2PLOT geometry into a VRML file, then importing the VRML file into
Acrobat 3D.  Within Acrobat 3D, the author can place the 3-d figure
anywhere within the PDF file, add annotations, and set default
rendering, lighting and viewing properties.  Where necessary, special
views that exhibit particular features of the figure discussed in the
text of the paper can be saved and named for the reader to select.

We now give three examples of relatively simple, 3-dimensional figures
that are illustrative of common plotting requirements in astronomy: a
dark matter merger tree, a survey catalogue redshift distribution and
a colour-magnitude diagram.

\subsection{Example 1: Halo merger trees}

Semi-analytic modelling has been developed as an efficient way to 
study the hierarchical formation of galaxies \citep[see e.g.][]{White91,
Cole00}.  The first stage in the process is to generate a merger tree of 
dark matter halos from direct $N$-body simulations or Monte Carlo 
methods. Analytic solutions are applied on top of the dark matter framework
to treat the baryonic component, and to model the physical processes required
to build a galaxy (initial mass function, star formation history, 
feedback, etc.).  While merger trees are predominantly studied statistically, 
individual trees feature intricate and exquisite structure that can 
only be fully appreciated via interactive 3-d visualisations.  Conventional,
static 2-d projections simply fail to represent this structure.

Figure~\ref{fig:merger} is a 3-d representation of a  merger tree, using 
data from the Virgo-Millennium 
Database\footnote{\url{http://www.g-vo.org/Millennium}}
\citep{Springel05,Delucia07}. Colour indicates the formation redshift for 
the progenitor haloes that combine through mergers to produce the final
galactic halo.  By rotating this merger tree, the filamentary
LSS within which the progenitors form is clearly visible.  
\begin{figure}
\rule{\textwidth}{0.2mm}
\vspace{0.05in}
\begin{center}
{\bf Please obtain paper with figures from: \newline
\url{http://astronomy.swin.edu.au/s2plot/3dpdf}}
\end{center}

\vspace{3in}

\rule{\textwidth}{0.2mm}
% XXX PUT COLOUR WEDGE ON LEFT OF 3-d FIGURE XXX
\caption{A merger tree showing the hierarchical formation of a galactic
dark matter halo, using data obtained from the Virgo-Millennium Run 
database.  Colour indicates the formation epoch of progenitor haloes.
\label{fig:merger}} 
\end{figure}

\subsection{Example 2: Redshift catalogues}

The redshift wedge or cone diagram is somewhat peculiar to astronomy.  
It is typically used to exhibit large-scale
structure (LSS) as (right ascension, declination, redshift)-tuples, and 
can comprises several individual plots to show
different slices from a larger volume of the Universe.  This is a
classic case of the traditional publishing medium (i.e. paper) limiting the
communication of genuine 3-dimensional information, and the
opportunity for advancement by provision of interactive 3-d figures is
clear.

In Figure~\ref{fig:hicat} we present an interactive, 3-dimensional
plot of the redshift distribution of the combined HICAT
\citep{Meyer04} and HICAT+N \citep{Wong06} galaxies.
\citet{Meyer04} presented two sky projections of HICAT in its entirety,
coloured by redshift, as well as two full-page figures for a
multi-layer wedge diagram.  Here, for the first time we combine HICAT
and HICAT+N into one figure, using the same colour coding as
\citet{Meyer04}.  The reader is invited to explore the plot by moving
the camera around, into and out of the galaxy distribution.  
The filamentary LSS is much more obvious than in a static figure,
and the local void is particularly noticeable from certain orientations.  
The coordinate grid and labels may be toggled (off) on by expanding the model tree 
and (de)selecting the {\tt GRID} and {\tt LABELS} nodes.
\begin{figure}
\rule{\textwidth}{0.2mm}

\begin{center}
{\bf Please obtain paper with figures from: \newline
\url{http://astronomy.swin.edu.au/s2plot/3dpdf}}
\end{center}

\vspace{3.4in}
{\tiny
\begin{minipage}{1in}
\begin{tabular}{clclcl}
\multicolumn{6}{l}{\bf Point colour} \\
\textcolor{hcyan}{$\bullet$} & $V_{\rm CMB} \leq 1000$ km s$^{-1}$ & \textcolor{hpink}{$\bullet$}  & $2000 < V_{\rm CMB} \leq 3000$ km s$^{-1}$ & \textcolor{hgreen}{$\bullet$}  & $4000 < V_{\rm CMB} \leq 5000$ km s$^{-1}$ \\
\textcolor{hblue}{$\bullet$} & $1000 < V_{\rm CMB} \leq 2000$ km s$^{-1}$ & \textcolor{hred}{$\bullet$}  & $3000 < V_{\rm CMB} \leq 4000$ km s$^{-1}$ & \textcolor{hyellow}{$\bullet$}  & $5000 < V_{\rm CMB}$ km s$^{-1}$ 
\end{tabular}
\end{minipage}
}

\rule{\textwidth}{0.2mm}
\caption{Interactive, 3-dimensional plot of the distribution of
galaxies in the combined HICAT and HICAT+N catalogues.\label{fig:hicat}}
\end{figure}

\subsection{Example 3: Colour-magnitude diagrams}

Widely-used 2-dimensional plots such as colour-magnitude
diagrams (CMDs) can benefit from a 3-dimensional treatment.  CMDs are
most frequently presented as scatter plots, where each
measurement is added to the plot as a dot.  The quantitative
interpretation of these plots is becoming more difficult as
star counts increase with instrument capability.  One obvious improvement is to
dispense with the scatter plot approach entirely.

In Figure~\ref{fig:cmd} we demonstrate that a simple 2-d histogram is
an effective replacement for a conventional CMD.  The source data for
the figure are 28\,568 stellar photometry measurements for the Large
Magellanic Cloud globular cluster NGC~1898 made with Hubble WFPC2
\citep{Olsen99}.  The figure shows the result of binning the $(M_V,
V-I)$ pairs onto a grid, and plotting the result as a logarithmic 2-d
histogram.  In certain orientations, the surface colour alone aids the
eye in interpreting relative star counts; viewed ``side-on'' the
height of the surface above the base gives a direct and quantitative
comparison between different parts of the colour-magnitude plane:
something that colour alone cannot accomplish.  Some choices of colour
scales, such as the standard ``rainbow'' map, should be avoided for
presenting relative data values - grey-scales offer a more intuitive
choice \citep{Tufte90}.
\begin{figure}
\rule{\textwidth}{0.2mm}

\begin{center}
{\bf Please obtain paper with figures from: \newline
\url{http://astronomy.swin.edu.au/s2plot/3dpdf}}
\end{center}

\vspace{3in}

\rule{\textwidth}{0.2mm}
\caption{2-dimensional colour-magnitude histogram of the stellar
population of NGC~1898.\label{fig:cmd}} 
\end{figure}

\section{S2PLOT, JavaScript and 3-d PDF}
\label{sec:js}

The JavaScript capabilities of 3-d PDF enable us to go well beyond the
relatively simple examples shown in
Figures~\ref{fig:hicat}--\ref{fig:merger}.  The S2PLOT graphics model
divides the scene into {\em static}\/ and {\em dynamic}\/ geometry.
The former is drawn once and never changes, while the latter can be
redrawn every refresh cycle and enables the construction of
visualisations that evolve with time or in response to user input.  By
embedding small JavaScript components in a PDF file, we can propagate
some of S2PLOT's dynamic features into 3-d PDF figures.

There are two basic reasons for using JavaScript in a PDF file: (i) to
offer the user {\em explicit}\/ control of the 3-d figure(s) in the
file, and (ii) to {\em implicitly}\/ modify the 3-d figure(s) in
response to internal events or user actions.  Adobe JavaScript for
Acrobat 3D is sufficiently complete that it is reasonable to expect to
see genuine, fully-fledged visualisation applications as components of
PDF files before long.  In the meantime, we describe two examples that
demonstrate how explicit and implicit JavaScript control can be used
to immediately enhance the use of 3-d figures in astronomy.

\subsection{Example 4: Large scale structure isosurfaces}

In this era of $N \sim10^6$ galaxy 
surveys \citep{Adelman07} and $N \sim 10^{10}$ particle simulations 
\citep{Springel05}, scatter plot wedge diagrams quickly become
overcrowded, and the large scale structures we aim to understand are
lost in a sea of points.  An alternative approach is to draw
isodensity contours (usually calculated on a regular mesh) over the
point distributions.  These help to highlight connected structures,
and isolate the regions of highest (or lowest) density.

In Figure~\ref{fig:isodensity}, isodensity surfaces are overlaid on a
dark-matter only cosmological simulation ($\Omega_{\rm M} = 0.24,
\Omega_{\Lambda} = 0.76, h = 0.73$, where cosmological parameters have
their usual meaning, and the simulation box length was 50 $h^{-1}$
Mpc).  While the full simulation had $256^3$ million particles, we
only plot $5\,000$ of these.  Particles are smoothed onto a $32^3$
mesh using a triangular shaped cloud kernel \citep{Efstathiou85}, and
the 40\%, 50\% and 60\% isodensity levels are generated.  By clicking
on the {\tt NEXT} button, it is possible to move between these levels,
demonstrating explicit JavaScript control.  The other buttons toggle
the display of the particle distribution and the bounding box.
\begin{figure}
\rule{\textwidth}{0.2mm}

\begin{center}
{\bf Please obtain paper with figures from: \newline
\url{http://astronomy.swin.edu.au/s2plot/3dpdf}}
\end{center}

\vspace{3.4in}
{\tiny
\begin{minipage}{1in}
\begin{tabular}{cl}
\multicolumn{2}{l}{\bf Surface colour} \\
\textcolor{hblue}{$\bullet$} & 40\% \\
\textcolor{hgreen}{$\bullet$}  & 50\% \\
\textcolor{hpink}{$\bullet$}  & 60\%
\end{tabular}
\end{minipage}
}

%\vspace{3in}
\rule{\textwidth}{0.2mm}
% XXX PUT COLOUR WEDGE ON LEFT OF 3-d FIGURE XXX
\caption{40\%, 50\% and 60\% isodensity levels from a dark matter-only
cosmological simulation.  Clicking the {\tt NEXT} button moves between
the three isodensity levels, demonstrating explicit JavaScript control
within a 3-d PDF figure.
\label{fig:isodensity}}
\end{figure}

\subsection{Example 5: Substructure in a dark matter halo}

The Cold Dark Matter (CDM) model, and its variants, has been very
successful at explaining a number of observational properties of
galaxies and galaxy clustering, particularly on large scales.
However, on small scales and/or in high density regions,
contradictions between CDM model predictions and observations
exist. One such example is the ``missing satellite'' problem
\citep{Bardeen96,Kauffmann93,Klypin99, Moore99,Kamionkowski00}: high
resolution CDM simulations of Milky Way-type dark matter haloes
produce many more bound sub-structures than there are observed
satellites.  Visualisation techniques are extremely valuable for
studying individual haloes, in order to better understand the role and
spatial distribution of sub-structures.

Figure~\ref{fig:substruct} shows a 3-d model of a dark matter halo
identified from within a CDM simulation using a friends-of-friends
algorithm.  The visualisation chosen is a direct volume rendering,
which in this case highlights regions of higher density.  In S2PLOT,
real-time volume rendering is achieved by creating three orthogonal
sets of slices (textures) through the data volume, then for every
redraw choosing and transparently layering the set that is most
orthogonal to the camera view direction.  While the transitions
between different texture sets are occasionally visible, this is a
very effective method for real-time inspection of a gridded volume.
When writing VRML, the three texture sets are exported and stored in
uniquely-named parts of the VRML model tree.  A simple JavaScript
attached to the 3-d figure in the PDF file is then used to determine
which texture set to display on every redraw cycle.  This JavaScript
runs ``behind the scenes'' to add implicit control over the rendering
in the 3-d PDF file.
\begin{figure}

\rule{\textwidth}{0.2mm}

\begin{center}
{\bf Please obtain paper with figures from: \newline
\url{http://astronomy.swin.edu.au/s2plot/3dpdf}}
\end{center}

\vspace{3in}
\rule{\textwidth}{0.2mm}
% XXX PUT COLOUR WEDGE ON LEFT OF 3-d FIGURE XXX
\caption{Volume rendering enhances the visibility of sub-structures in
a dark matter halo, compared to only plotting locations of points.  
By querying the location of the camera relative to the model, implicit 
JavaScript control determines which of the three sets of volume rendered 
textures should be displayed.
\label{fig:substruct}} 
\end{figure}

\section{Other approaches}
\label{sec:other}

While the preceding examples have highlighted the data visualisation
advantages of 3-d PDF, there are other aspects of academic publication
where the CAD/CAM functionality can be used.  Two such cases are in
instrument design and the presentation of instructional models (3-d
cartoons), which we now demonstrate.

\subsection{Example 6: Instrument design}

Advances in astronomy are intimately linked to technological
developments, particularly through new instrument designs which
enhance the capability of existing facilities.  CAD/CAM packages are a
natural source of engineering models for astronomy instrumentation,
but a range of polygon modelling packages exist, many developed and
used within the computer animation industry.

As a means of promoting design concepts for the Square Kilometer Array
(SKA) and Australian SKA Pathfinder (ASKAP), one of us (Fluke) worked
with engineers from the CSIRO Australian National Telescope Facility
to create a series of 3-d radiotelescope models.\footnote{3DS human
model from \url{http://www.klicker.de}} Based loosely on engineering
drawings, the telescopes were built using NewTek's Lightwave 3D
V8.0,\footnote{\url{http://www.newtek.com}} and exported for Acrobat
3D in 3DS format. This binary format stores geometrical objects as a
mesh of 3-vertex polygons with associated materials (i.e. colours).
The 3DS format can be imported without further conversion to Acrobat
3D, and the model tree allows selection of individual components based
on the material.  The result is shown in Figure~\ref{fig:telescope}.
\begin{figure}
\rule{\textwidth}{0.2mm}

\begin{center}
{\bf Please obtain paper with figures from: \newline
\url{http://astronomy.swin.edu.au/s2plot/3dpdf}}
\end{center}

\vspace{3in}
\rule{\textwidth}{0.2mm}
% XXX PUT COLOUR WEDGE ON LEFT OF 3-d FIGURE XXX
\caption{A model radiotelescope constructed as a part of a promotional 
campaign for the Australian SKA Pathfinder. By opening the model tree, 
individual components of the model may be (de)selected. This is an example
of 3-d PDF's original raison detre.
\label{fig:telescope}} 
\end{figure}

\subsection{Example 7: Instructional diagrams}

An important component of many research articles and textbooks is the
instructional diagram, often presented in a simplified cartoon format.
These diagrams usually show spatial and/or other relationships between
the elements of a model.  In many cases only a single orientation of a
model is presented yet, as with astronomical datasets, these models
are often inherently three-dimensional.  3-d PDF makes possible the
inclusion of simple but interactive, 3-d diagrams in electronic
articles, with applications in teaching, pedagogy and outreach.

One particular instructional diagram common in astronomy is the
unified model for Active Galactic Nuclei (AGN) and quasars
\citep[e.g.][]{Antonucci93}.  Classification of an AGN as either Type
1, Type 2 or Blazar depends on the orientation of the dusty molecular
torus to the observer's line-of-sight. The viewing angle determines
the presence or absence of broad and narrow line emission, and beamed
radiation from jets.  In Figure~\ref{fig:agn} we present an
interactive 3-d AGN (components of the model are not to scale). By
rotating the model, the reader is able to ``observe'' the AGN from
arbitrary viewpoints, and see for example the obscuration of the broad
line region and central engine by the torus.
\begin{figure}
\rule{\textwidth}{0.2mm}

\begin{center}
{\bf Please obtain paper with figures from: \newline
\url{http://astronomy.swin.edu.au/s2plot/3dpdf}}
\end{center}

\vspace{3in}
\rule{\textwidth}{0.2mm}
% XXX PUT COLOUR WEDGE ON LEFT OF 3-d FIGURE XXX
\caption{The unified model of Active Galactic Nuclei presented in 
interactive form (components are not strictly to scale).  The viewing
angle determines the visibility of the broad line region-emitting
``clouds'' surrounding the central engine, and the orientation of the
jets.  Presented in this form, the view-dependent classification of
AGN type may for some people be more instructional than in a series of
static 2-d images.
\label{fig:agn}} 
\end{figure}

\section{Closing Remarks}
\label{sec:conc}

We have demonstrated that the new 3-d extensions to the PDF standard
can be used to publish instructive and interactive 3-d figures in
astronomy research papers.  Our approach uses the S2PLOT programming
library and VRML as an intermediary 3-d model description format.  It
is sufficiently easy to create and embed 3-d figures in PDF articles
that we believe it can become a standard technique in the publication
of scientific research results, particularly in the fields of
astronomy and astrophysics.  3-d PDF files are portable, relatively
compact and viewable on many desktop platforms today.\footnote{Adobe
Reader 8 is available for Microsoft Windows and Apple Mac OS X systems
now, versions for Linux and other major unix platforms are due
shortly.}  Small changes to the publishing workflow are required, but
we contend they are insignificant in the context of the major
advantages 3-dimensional figures and diagrams will bring to inter- and
intra-domain science communication.

There is one major caution to be made on the use of 3-d PDF in
science.  Scientists traditionally use large collections of {\em
points}\/ to representing instances in parameter spaces, while the
CAD/CAM community works primarily with {\em surfaces}.  Like their
CAD/CAM and animation software counterparts, Acrobat 3D and Reader
show a clear preference for mesh (surface) data over point data.  We
found that the 3-d PDF reader client (Acrobat Reader) --- and to a
lesser extent the content creation tool (Acrobat 3D) --- did not
perform well when more than $\sim 10$--20\,000 points were used in a
single 3-d figure. Gridding point data and then rendering isosurfaces
and/or volumes is one effective work-around.  Some improvement in the
underlying software is required before 3-d figures of eg.\ redshift
catalogues containing $10^5$ to $10^8$ galaxies can be published,
although doing so would have an obvious impact on the file size of the
article.

In this paper we have only touched on the possibilities of 3-d PDF for
astronomy and astrophysics.  We envisage many further applications of
the technology, but specifically we are interested in further
enhancing the interaction the user can have with the figure.  For
example, it is possible with JavaScript to capture mouse clicks on
individual elements of the 3-d scene; this could be used to directly
select a galaxy (or galaxies) in a redshift catalogue visualisation
and report their names to the user.  In principle, object names
gathered this way could then be used in JavaScript queries to remote
data sources such as virtual observatory services.  Additional
proposals include developing uses for the Reader's measuring tool, and
if possible, calculating and reporting statistics on regions selected
by the user.

Finally, the opportunity to improve the science community's capacity
to convey advances in our disciplines to the wider public should not
be ignored.  PDF is the most widely-used, self-contained electronic
document format.  Our funding agencies, governments and public can now
easily interact with instructive, 3-dimensional graphical
representations of our work --- all we have to do is create and
publish them.

\section*{Acknowledgements}

We thank Alyssa Goodman and colleagues at the Initiative in Innovative
Computing at Harvard for discussions on their approach to 3-d PDF. We are
grateful to Chris Power for providing cosmological simulation data.
The telescope model shown was originally developed with Peter Hall and
colleagues at the CSIRO Australia Telescope National Facility.  Barnes
acknowledges the support of iVEC visitor fund which enabled parts of
this work.

% \appendix

% \section{}
% \label{}

% Bibliographic references with the natbib package:
% Parenthetical: \citep{Bai92} produces (Bailyn 1992).
% Textual: \citet{Bai95} produces Bailyn et al. (1995).
% An affix and part of a reference:
%   \citep[e.g.][Ch. 2]{Bar76}
%   produces (e.g. Barnes et al. 1976, Ch. 2).

\end{document}